\documentstyle[12pt]{article}

\parskip=\medskipamount % space between paragraphs (LaTeX)
\def\be{\begin{equation}}
\def\bea{\begin{eqnarray}}
\def\nn{\nonumber}
\def\ee{\end{equation}}
\def\eea{\end{eqnarray}}
\def\CR{\hbox{{$\cal R$}}} 
 
 \def\CP{\hbox{{$\cal P$}}}

\def\sgn{{\rm sgn}}

\begin{document}
\begin{titlepage}

\begin{center}
{\Large \bf Anyons as  quon particles}\\
\vspace{2.cm}
{\normalsize \bf V.Bardek, M.Dore\v si\' c, and S.Meljanac\footnote
{e-mail address: bardek@thphys.irb.hr\\
 e-mail address: doresic@thphys.irb.hr\\
 e-mail address: meljanac@thphys.irb.hr}}\\
\vspace{0.5cm}
Department of Theoretical Physics, \\
Rudjer Bo\v skovi\' c Institute, P.O.B. 1016,\\
41001 Zagreb,CROATIA \\
\vspace{2cm}
{\large \bf Abstract}
\end{center}
\vspace{0.5cm}
\baselineskip=24pt
The momentum operator representation of nonrelativistic anyons is
developed in the Chern - Simons formulation of fractional statistics.
The connection between anyons and the q-deformed bosonic algebra is
established.
\vspace{70mm}
\begin{center}
PACS numbers:~03.65.-w,03.70.+k,05.30.-d,11.10.Ln.
\end{center}
\end{titlepage}

\setcounter{page}{1}
\newpage
\baselineskip=24pt
\vspace{1.5cm}     % ( 1 in = 2.54 cm )
It is a well-known phenomenon that particles in ~ 2+1 ~ dimensions
 minimally coupled to an Abelian gauge field with the Chern - Simons
 Lagrangian change their spin and statistics ~ \cite{wil}.The
  corresponding vector potential can be absorbed in a new matter
  field that obeys graded , anyonic commutation relations
 ~ \cite{jack}.Commutators are changed into anticommutators for
special values of the Chern - Simons parameter.For other values ,the
interpolating anyon statistics appears.This is the essence of the
so-called statistical transmutation.

Floratos and Tomaras ~ \cite{flor} have recently proposed a connection
between the q - deformed harmonic oscillator and the discretized
two-anyon systems, with  q being a primitive root of unity and
 the relative radial motion of anyons frozen.

On the other hand,Greenberg ~ \cite{gr} has shown that the
q-deformation ( q being a real parameter )of the bilinear Bose
(or Fermi) commutation algebra introduces particles called "quons"
that interpolate between fermions and bosons.

In this paper we develop a momentum-operator representation of
nonrelativistic anyons.We demonstrate that anyons can be alternatively
represented as a q-deformation of an underlying bosonic algebra.This
 can be viewed as an extension of Greenberg's approach with q being a
 complex number, $\mid q \mid = 1 $.
Furthermore,we construct ~$\CR$~-matrices for anyon operators both
in position and momentum space and show that they satisfy the
Yang-Baxter equation and the Hecke condition.

The quantum-mechanical problem of nonrelativistic bosons interacting
with the U(1) gauge field described by the Chern-Simons Lagrangian has
 been formulated and solved by Jackiw and Pi ~ \cite{pi}.They have
 shown that the corresponding Hamiltonian can be reduced to the
 following form:
\begin{eqnarray}
 H & = & \frac{1}{2} \int d^2 r \; \vec {\bf {\sqcap}}^{\dagger}
 ({\vec r} \, ) \; \vec {\bf {\sqcap}} ({\vec r} \, ) \: .  \label{1}
\end{eqnarray}
The operator covariant derivative $\vec {\bf{\sqcap}}$ is given by
\begin{eqnarray}
 \vec {\bf {\sqcap}} \, ({\vec r} \, ) & = & ( \vec \nabla - i \:
 \vec A( \vec r \, )) \: \Psi( \vec r \, ) \; .             \label{2}
\end{eqnarray}
The quantum-field operator  $\Psi( \vec r \, )$ and its Hermitian
conjugate $\Psi^{\dagger}( \vec r \, )$ obey the bosonic commutation
relations at equal times:
\begin{eqnarray}
[ \; \Psi ( \vec r \, ) \; , \; \Psi ( \vec r \, ' \, ) \; ] &=& 0
\; ,
 \nonumber\\
{[ \; {\Psi}^{\dagger} ( \vec r \, ) \; , \; {\Psi}^{\dagger}
( \vec r \, ' \, ) \; ]} &=& {0 \; ,}
 \nonumber\\
{[ \; \Psi( \vec r \, ) \; , \; {\Psi}^{\dagger} ( \vec r \, ' \, )
 \; ]}
 &=& { \delta \, ( \, \vec r - \vec r \, ' \, ) \; .}       \label{3}
\end{eqnarray}
The vector potential ~$\vec A$~ can be completely described by the
bosonic number density operator ~$\rho$ ~ :
\begin{equation}
\vec A \: ( \vec r \, ) \; = \; - \lambda \: \hat{n}  \: {\times} \:
 {\int} \frac{ \vec r - \vec r \, '}{{\mid \vec r - \vec r \, '
 \mid}^2} \; \rho( \vec r \, ' \, ) \; d^2 r' \; ,          \label{4}
\end{equation}
\begin{equation}
\rho( \vec r \,) \; = \; {\Psi}^{\dagger}({\vec r} \, ) \:
\Psi ({\vec r} \, ) \; ,                                    \label{5}
\end{equation}
where ~ $\lambda$ ~ is the so-called statistical parameter whose role
will be clarified below.In other words,the gauge field has no
independent dynamics.In this way,the vector potential is completely
eliminated at the expense of the Hamiltonian becoming nonlocal.\\
As has been noted in ~\cite{for},the definition of~ $\vec A$ ~ as a
pure gauge is
\begin{equation}
\vec A \: ( \vec r \, ) \; = \; \vec \nabla \, \omega
( \vec r \, ) \; ,                                          \label{6}
\end{equation}
where
\begin{equation}
\omega ( \vec r \, ) \; = \; - \lambda \int \theta
( \, \vec r - \vec r \, ' \, ) \rho ( \vec r \, ' \, ) \; d^2 r' \; .
                                                            \label{7}
\end{equation}
This definition enables one to transform the nonlocal interaction
 away.However, the polar angel ~$\theta$ ~ has to be a multivalued
 function; otherwise there would be residual interactions
 originating from the cut.By the phase redefinition of the bosonic
 field,
\begin{equation}
\Psi( \vec r \, ) \; = \; e^{i \, \omega ( \vec r \, )} \:
\tilde{\Psi} ( \vec r \, ) \; ,                             \label{8}
\end{equation}
we obtain the following free Hamiltonian of the new anyonic field
$\tilde{\Psi} \; : $
\begin{equation}
\tilde{H} \; = \; \frac{1}{2} \int d^2 r \; \vec \nabla
{\tilde{\Psi}}^{\dagger} ( \vec r \, )
\vec \nabla \tilde{\Psi} ( \vec r \, ) \; .                 \label{9}
\end{equation}
The commutation relations~(3)~ are now modified as
\begin{eqnarray}
\tilde{\Psi}(\vec r \, ) \: \tilde{\Psi}(\vec r \, ' \, )
\; - \; e^{-i \, \lambda \, \Delta( \vec r - \vec r \, ' \, )}
\tilde{\Psi}(\vec r \, ' \, ) \: \tilde{\Psi}(\vec r \, )
&=& 0 \; ,
\nonumber\\
\tilde{\Psi}^{\dagger}(\vec r \, ) \: \tilde{\Psi}^{\dagger}
(\vec r \, ' \, )
\; - \; e^{-i \, \lambda \, \Delta( \vec r - \vec r \, ' \, )}
\tilde{\Psi}^{\dagger}(\vec r \, ' \, ) \: \tilde{\Psi}^{\dagger}
(\vec r \, )
&=& 0 \; ,                                                  \label{10}
\\
\tilde{\Psi}(\vec r \, ) \: \tilde{\Psi}^{\dagger}(\vec r \, ' \, )
\; - \; e^{i \, \lambda \, \Delta( \vec r - \vec r \, ' \, )}
\tilde{\Psi}^{\dagger}(\vec r \, ' \, ) \: \tilde{\Psi}(\vec r \, )
 & = & \delta \,( \, \vec r - \vec r \, ' \, ) \; .
\nonumber
\end{eqnarray}
Here ~ $\Delta$~ denotes the difference
\begin{equation}
\Delta \,( \, \vec r - \vec r \, ' \, ) \; = \;
\theta \, ( \, \vec r - \vec r \,' \, ) \; - \;
\theta \, ( \, \vec r \, ' - \vec r \, ) \; .               \label{11}
\end{equation}
Because of the multivaluedness of ~$\theta$~ , this difference can be
identified with the infinite set ~$\{ \, \pi \, + \, 2 \pi z \, \}
\; ,$ ~z ~being an integer.
When ~$\lambda$~ takes a rational value, the multivaluedness is finite.
The multivalued nature of the phase factor
 ~$\Delta$~ is essential to the consistency of the graded commutation
  relations ~(10).The excitations described by the new field ~$\tilde
   \Psi$~ obey fractional statistics.\\ For ~$\lambda \, = \, 0 \,
    mod \, 2 \; ,$ we have bosons; for ~$\lambda \, = \, 1 \, mod \,
     2 \; ,$ we have fermions , whereas for an arbitrary value of
     ~$\lambda \, ,$~ we have anyons.The graded commutators~(10)~
     resemble the q-deformed algebras considered by Greenberg~[4].
Note,however,that this algebra was formulated in momentum space.
In order to compare and further analyze such algebras,one should be
able to evaluate a full quantum-field theoretic description of anyons
in momentum space.Hence,we should Fourier transform the anyonic fields
and the graded
commutation relations in Eq.~(10).However,the problem is how to define
 the integration  of a multivalued function precisely;for example ,
 in Eq.~(7).

We define the integration of a multivalued function as a
set of integrals of single-valued functions defined in the
correspoding sheets of the Riemann surface.This set of single-valued
functions we then treat as one multivalued function.

Now we apply this definition to the multivalued angle function
{}~$\theta (\vec r \, )$~.In the zeroth sheet, we define polar
angle as ~${\theta}_{0}(\vec r \, )$~with
{}~${\theta}_{0}(\vec r \, ) \in [0,2 \pi ) $~ which satisfies
\be
{\theta}_{0}(\vec r \, ) - {\theta}_{0}(- \vec r \, ) =
{\Delta}_{0}(\vec r \, ) \: ,                               \label{21}
\ee
where
\bea
{\Delta}_{0}( \, \vec r \, ) &=&
    \left \{ \begin{array}{ll}
             - \pi \: \sgn \, y(\vec r \, )  & {\rm if} \; \;
             y \neq 0  \\
             - \pi \: \sgn \, x(\vec r \, )  & {\rm if} \; \;
             y \, = \, 0
             \; .                                           \label{22}
             \end{array}
    \right.
\eea
The multivalued angle function is
\bea
\theta(\vec r \, ) &=& {\theta}_{0}(\vec r \, ) + 2 \pi z \, ,
\; \; \; z \in Z \: ,
\nn \\
\theta(- \vec r \, ) &=& {\theta}_{0}(- \vec r \, ) +
{\Delta}_{0}(\vec r \, ) - \Delta \: ,                      \label{23}
\eea
satisfying
\be
\theta(\vec r \, ) - \theta(- \vec r \, ) =
\Delta = \pi + 2 \pi z \: , \; z \in Z \: .                 \label{24}
\ee
The Fourier transform of the multivalued function is calculated
according to (using) our definition of the integration
of the multivalued function.For example, let us consider multivalued
constant ~$\Delta \equiv  \pi + 2 \pi z , \: z \in Z$~.
In the n-th sheet,
the value is ~$\pi + 2 \pi n$~ and its Fourier transform is
{}~$( \pi + 2 \pi \, n ) \delta (\vec k \, )$~.
Hence the Fourier transform of the multivalued constant in
momentum space is~${( \pi + 2 \pi z )} \delta (\vec k \, )$~.

Let us now construct annihilation and creation anyon operators in
 momentum space by Fourier transforming the anyon field
 ~$\tilde \Psi$~ and its
Hermitian conjugate ~$\tilde {\Psi}^{\dagger} \; :$~
\begin{eqnarray}
\tilde a( \vec k \,) &=& \frac{1}{2 \pi} \int e^{-i \vec k \vec r}
\tilde \Psi( \vec r \, ) d^2 r \; ,
\nonumber\\
\tilde a^{\dagger}( \vec k \,) &=&
\frac{1}{2 \pi} \int e^{i \vec k \vec r}
\tilde {\Psi}^{\dagger}( \vec r \, ) d^2 r \; .             \label{13}
\end{eqnarray}
In terms of the above operators, the commutator algebra~(10)~
 translates into
\begin{eqnarray}
\tilde a( \vec p \, ) \tilde a( \vec q \, ) \; - \;
 \frac{1}{4 \pi^2} \int d^2 k \int d^2 r \;
e^{-i \lambda \Delta( \vec r \, )+i \vec k \vec r} \;
\tilde a( \vec q - \vec k \, ) \tilde a( \vec p + \vec k \, )
 &=& 0 \; ,
\nonumber\\
\tilde a^{\dagger}( \vec p \, ) \tilde a^{\dagger}( \vec q \, )
 \; - \; \frac{1}{4 \pi^2} \int d^2 k \int d^2 r
e^{-i \lambda \Delta( \vec r \, ) + i \vec k \vec r} \;
\tilde a^{\dagger}( \vec q + \vec k \, ) \tilde a^{\dagger}
( \vec p - \vec k \, )  &=& 0 \; ,                          \label{14}
\\
\tilde a( \vec p \, ) \tilde a^{\dagger}( \vec q \, ) \; - \;
\frac{1}{4 \pi^2} \int d^2 k \int d^2 r
e^{i \lambda \Delta( \vec r \, ) + i \vec k \vec r} \;
\tilde a^{\dagger}( \vec q + \vec k \, ) \tilde a ( \vec p +
\vec k \, ) &=& \delta( \vec p - \vec q ) \; .
\nonumber
\end{eqnarray}
For anyonic values of the statistical parameter ~$\lambda$~, the
 graded commutation relations ~(\ref{14})~ imply the hard-core
condition on one-particle operators:
\begin{eqnarray}
\tilde a^2 ( \vec p \,) \; = 0 \; \; \; , \; \; \;
 {\tilde a}^{\dagger \: 2} ( \vec p \,) \; = 0 \; .         \label{15}
\end{eqnarray}
{}From relation ~(11)~ we know that ~$\Delta$~ is actually a
multivalued constant and
can therefore be removed outside the integration in ~(\ref{14})~.The
resulting algebra has the same form as the algebra in position space
{}~(10)~.It is obvious that the connection with a q-deformed algebra
is a direct one.We simply identify the q-parameter and the
multivalued unimodular complex number ~$e^{i \lambda \pi (1+2z)}$~.
This is,in some sense, a generalization of the usual (standard) notion
of q-deformed field theory considered by Greenberg.By substituting
the Fourier decomposition of the anyon fields ~$\tilde \Psi$
{}~~(\ref{13})~
into the free Hamiltonian ~(9)~,we can easily obtain the corresponding
Hamiltonian in momentum space:
\begin{equation}
\tilde H \; = \; \frac{1}{2} \int d^2 p \: {\vec p}^{\, 2} \:
\tilde a^{\dagger}( \vec p \,) \tilde a( \vec p \,) \; ,    \label{16}
\end{equation}
where ~${\vec p}^{\: 2} / 2$~ is the single-particle energy for a
noninteracting system.The Hamiltonian operates in the Fock space
generated by repeated application of the creation operators
{}~$\tilde a^{\dagger} ( \vec p \, )$~ to the vacuum ~$\mid 0>$~,which
is,by definition, annihilated by all  the operators
{}~$\tilde a ( \vec p \, )$~:
\begin{eqnarray}
\tilde a ( \vec p \, ) \mid 0> \; = \; 0 \; \; \; for \; all \; \;
\vec p \; .                                                 \label{17}
\end{eqnarray}
Using the commutation algebra ~(\ref{14})~ and the hard-core condition
{}~(\ref{15})~,it is easily verified that all states in our momentum space
 have a non-negative squared norm.\\
One can use the bosonic (q=1) operators ~$a( \vec k \,)$~ as building
blocks to construct a momentum representation of the anyon operator
in the general case\\
{}~$( \;q \; = \; e^{i \lambda \pi (1+2z)} \;
\neq \; 1 \; )$~.By Fourier transforming relation ~(8)~,we obtain
\begin{eqnarray}
\tilde a( \vec k \, ) & = & a( \vec k \, ) + i \lambda
\int \: d^2 k' \; \theta( \vec {k'} \, ) \rho ( \vec {k'} \,)
a( \vec k - \vec {k'} \, ) \;
\nonumber\\
 & & - \frac{{\lambda}^2}{2} \int d^2 k' d^2 k'' \;
 \theta( \vec {k'} \, )
 \theta( \vec {k''} \, ) \rho ( \vec {k'} \,) \rho ( \vec {k''} \,)
a( \vec k - \vec {k'} - \vec {k''} \, ) + \dots \; \; ,     \label{18}
\end{eqnarray}
where we have formally expanded the exponential kernel in powers of
the statistical parameter ~$\lambda$~.~$\rho ( \vec k \, )$~ denotes
the Fourier transform of the bosonic density ~$\rho ( \vec r \, )$~,
 and ~$\theta ( \vec k \, )$~ is given by
\begin{eqnarray}
\theta ( \vec k \, ) &=& \int d^2 r \: e^{-i \vec k \vec r } \,
\theta ( \vec r \, )
\nonumber\\
 &=&  4 \pi^3 ( 1 + 2z ) \delta ( \vec k \, ) \, + \,
 \frac{2 \pi}{k^2} \,
 \tan \alpha \, +  \,  2 \pi^2 i \delta ( k_x )  \CP
 \frac{1}{ k_y} \; \; ,                                     \label{19}
\end{eqnarray}
where ~$\alpha$~ is the polar angel of the vector ~$\vec k$~
and ~$\CP$~ means the principal-value distribution.
Note that
\bea
\theta(\vec k \, ) &=& {\theta}_{0}(\vec k \, )
+ 2 \pi z \, \delta(\vec k \, ) \: ,
\nn \\
\theta(- \vec k \, ) &=& {\theta}_{0}(- \vec k \, ) +
{\Delta}_{0}(\vec k \, ) - \Delta \,
\delta(\vec k \, ) \: .                                     \label{25}
\eea
It is easy to see that
 ~$\theta ( \vec k \, )$~ respects the consistency condition
\begin{eqnarray}
\theta ( \vec k \, ) \: - \: \theta (- \vec k \, ) &=&
(2z+1) \pi \delta ( \vec k \, ) \; ,                        \label{20}
\end{eqnarray}
which can be found by Fourier transforming relation ~(11)~ .
As far as the resemblance between the Greenberg model ~ \cite{gr}~and
 our multivalued picture of anyons is concerned, the latter does not
 contain any conjecture regarding the form of the representation
 ~(\ref{18})~.
This representation follows directly from the gauge transformation
{}~(8)~ which is the cornerstone of the Chern-Simons formulation of the
fractional statistics in ~ 2+1~ dimension.
On the other hand , Greenberg's approach is more general ,since
it does not depend on the dimensionality of the underlying space.

We point out that the q-commutator algebras both in position
space, Eq.~(\ref{10})~, and in momentum space, Eq.~(\ref{14})~,
are the same and
can be written in the ~$\CR$~-matrix approach ~\cite{zach}~.For example,
in position space we have
\bea
\tilde{\Psi}( \vec r_{1} \, ) \tilde {\Psi}( \vec r_{2} \, ) \; - \;
\int d^2 x \int d^2 y \;
{\cal R} ( \vec r_{1}, \vec r_{2}, \vec x , \vec y \, ) \:
\tilde{\Psi}( \vec y \, ) \tilde{\Psi}( \vec x \, )
 &=& 0 \; ,
\nn \\
\tilde {\Psi}( \vec r_{1} \, ) \tilde {\Psi}^{\dagger}
( \vec r_{2} \, ) \; - \;
\int d^2 x \int d^2 y \;
{\cal R} ( \vec y , \vec r_{1}, \vec r_{2}, \vec x \, ) \:
\tilde{\Psi}^{\dagger}( \vec y \, ) \tilde{\Psi}( \vec x \, )
&=& \delta ( \vec r_{1} - \vec r_{2} ) \; .                  \label{26}
\eea
The ~$\cal R$~-matrix is given by
\be
{\cal R} ( \vec r_{1}, \vec r_{2}, \vec r_{3}, \vec r_{4} \, ) \: = \:
e^{-i \lambda \Delta} \:
\delta ( \vec r_{1} - \vec r_{3} ) \,
\delta ( \vec r_{2} - \vec r_{4} ) \: ,                     \label{27}
\ee
with multivalued matrix elements.Note that all quantites in a
multivalud picture have a common multivalued unimodular factor
{}~${ e^{i 2 \pi \lambda z} \, , \, z \in Z }$~,which has the property
of unit element.

In order that the above ~$\cal R$~-matrix algebra be associative,
the following conditions have to be satisfied:

(i) the Yang-Baxter equation
\bea
&&\int \int \int d^2 x  d^2 y d^2 z \;
{\cal R} ( \vec r_{1}, \vec r_{2}, \vec x , \vec y \, ) \,
{\cal R} ( \vec y , \vec z , \vec r_{3}, \vec r_{4} \, ) \,
{\cal R} ( \vec x , \vec r_{5}, \vec r_{6}, \vec z \, ) \,
\nn \\
&& \; \; \; = \, \int \int \int d^2 x  d^2 y d^2 z \;
{\cal R} ( \vec r_{2}, \vec r_{5}, \vec x , \vec y \, ) \,
{\cal R} ( \vec z , \vec x , \vec r_{6}, \vec r_{3} \, ) \,
{\cal R} ( \vec r_{1} , \vec y , \vec z, \vec r_{4} \, ) \; ,
                                                            \label{28}
\eea
(ii) hermiticity
\be
e^{-i \lambda \Delta} \: = \: e^{i \lambda \Delta} \: ,     \label{29}
\ee
(iii) the Hecke condition
\bea
( \check{\cal R} \, - \, 1 ) \, ( \check{\cal R} \, + \, 1 )
\: = \: 0 \, ,                                              \label{30}
\eea
where
\bea
\check{\cal R} \: &=& \: P \, {\cal R} \; \; \; \; \; \; \; and
\nn \\
P ( \vec r_{1}, \vec r_{2},\vec r_{3}, \vec r_{4} \, ) \: &=& \:
\delta ( \vec r_{1} - \vec r_{4} \, ) \:
\delta ( \vec r_{2} - \vec r_{3} \, ) \; .                  \label{31}
\eea
It is straightforward to show that the ~$\cal R$~-matrix
satisfies all the above
conditions.Hence this proves the braiding properties
of anyonic operators.

Finally,let us briefly comment on the clustering decomposition
properties
of the vacuum matrix elements of products of free-anyon multivalued
fields.Like quon theory,anyonic physics in the Chern-Simons
description is
basically a nonlocal theory.
Nevertheless,clustering holds for our nonrelativistic anyon theory.
This follows from a similar type of arguments as in Ref.
 \cite{gr}~ using the algebra ~(10)~.The two-point function has to
be replaced by ~$\delta ( \vec r - \vec r  \, ')$~and the quon
parameter q identified by the multivalued parameter
{}~$e^{i \lambda \Delta}$~.

The next step is to study the physics of anyons in the single-valued
picture and,in particular,the effect of cut-dependent terms upon the
established q-deformed algebra of anyonic operators~(\ref{14})~.Further
study of this problem will be reported elsewhere.
{\newpage \large \bf Acknowledgment} \\
\vspace{1cm}
This work was supported by the Scientific Fund of the Republic of
Croatia.

\end{document}